**Evaluation of defect density by top-view large scale AFM on metamorphic structures grown by MOVPE**

Agnieszka Gocalinska,[1,*] Marina Manganaro,[1] Valeria Dimastrodonato,[1] and Emanuele Pelucchi[1]

*1: Tyndall National Institute, "Lee Maltings", University College Cork, Cork, Ireland*

## Abstract

We demonstrate an atomic force microscopy based method for estimation of defect density by identification of threading dislocations on a non-flat surface resulting from metamorphic growth. The discussed technique can be applied as an everyday evaluation tool for the quality of epitaxial structures and allow for cost reduction, as it lessens the amount of the transmission electron microscopy analysis required at the early stages of projects. Metamorphic structures with low surface defectivities (below $10^6$) were developed successfully with the application of the technique, proving its usefulness in process optimisation.

**Keywords**: Metalorganic vapor phase epitaxy, Atomic force microscopy, Defects, Arsenates, Semiconducting III-V materials

## Introduction

Defect density is one of the crucial parameters to be controlled in fabrication of high performance and temperature stable devices, e.g. for photovoltaics [1] or when they need to be operated at high temperatures [2]. Metamorphic buffer layers (designed to step- or gradually change the alloy composition) are becoming a popular solution to reach the desired in-plane lattice parameter in case of growths intended to be pseudomorphic, but not lattice matched to any of the market available epitaxial substrates. [3,4,5,6,7] In that case, the detection of dislocation on the surface is a challenging task due to the complex morphology induced by step-bunching and residual strain [8], which make surface features not easily identifiable [9] and growth technique dependent.

Cross-sectional transmission electron microscopy (TEM) measurements are often very limited in range to allow for reliable defect counting [10], and planar view TEM, which would be an ideal tool, is very costly and time consuming, preventing it from being a standard, every day laboratory measurement to investigate sample quality.

Atomic Force Microscopy (AFM) provides reproducible, non-destructive and cheap (in terms of instrument purchase cost, consumables and time necessary to obtain the image, in respect to, for example, TEM) way of imaging the surface. Very high resolution (limited basically by the tip sharpness) and large (tens on microns) scanning range allow for adaptable application to various morphologies, permitting to capture simultaneously small features and long scale organisation. Wide range of materials can be investigated by AFM, partially thanks to the tapping mode operation, which overcomes problems associated with friction, adhesion, electrostatic forces, etc. Moreover the instrument is easy to use, therefore is often present in even moderately large laboratories. Despite all those obvious advantages, surprisingly little information (namely figures presenting real AFM images, depicting exhaustively actual surface appearance with all its visual characteristics) can be found in literature regarding the surface morphology of planar structures, as authors usually limit themselves to quoting RMS only, in which case information about e.g. short and long range step organisation is lost.

[*] Corresponding author : agnieszka.gocalinska@tyndall.ie

Here we show how to identify threading dislocations and estimate defect density on the surfaces of metamorphic structures grown by metalorganic vapour phase epitaxy (MOVPE). Despite a complex topography, the defect lines can be visualised, and multiple, large scale imaging provides the necessary statistics for reliable defect counting. We discuss how the method can reduce the number of TEM scans on the early stages of structure development, when often multiple samples are grown, therefore to reduce the cost and increase the pace of the optimisation. Also, we demonstrate that it is necessary to use AFM as a complementary technique for the correct evaluation of surface quality to compensate for the limitations of other microscopy techniques, and we would like to stress here the importance of routine checking and reporting on the AFM morphologies at various stages of development of multistack/metamorphic structures.

**Material and Methods**

All epitaxial samples discussed here were grown in a high purity MOVPE [11] commercial horizontal reactor (AIX 200) at low pressure (80 mbar) with purified $N_2$ as carrier gas. The precursors were trimethylindium (TMIn), trimethylgallium (TMGa), arsine ($AsH_3$) and phosphine ($PH_3$). The sample designs and growth condition varied, as described in text and summarised in Table 1 [12]. $In_xGa_{1-x}As$ graded buffers were grown on GaAs substrate with gradually increased In concentration from below 1% (value corresponding to minimum controllable In flow allowed by the MOVPE system). The final reached concentration and the actual Ga-In exchange curves varied between samples and are described in more detail in Table 1. All samples had a homoepitaxial GaAs or InP 200 nm thick buffer grown prior to the graded $In_xGa_{1-x}As$. All layers were nominally undoped.

All epitaxial growths described resulted in relatively smooth surfaces with cross-hatch pattern clearly visible when inspected with an optical microscope in (Nomarski) differential interference contrast (N-DIC) or in dark field mode (not shown). Subsequent detailed morphological studies were performed with Veeco Multimode V Atomic Force Microscope in tapping/non-contact mode at room temperature and in air. Samples were scanned perpendicular to the cleaving edge, unless stated otherwise in the text (GaAs cleaves perpendicularly along (110) planes [13]). The defects formation and propagation was cross-correlated with cross sectional TEM when possible.

The assessment of composition and strain in the layers was made according to measurements of Reciprocal Space Maps (RSM) obtained by high resolution X-ray diffraction measurements (HRXRD). Measurements were done in a symmetric (004) and two asymmetric ((224) and (-2-24)) reflections with sample positioned at 0º, 90°, 180° and 270° with respect to its main crystallographic axes. [14]

## Table 1 Growth conditions, design and characterisation of all discussed MBLs.

| Figure | Growth number | Substrate misorientation | Growth temperature | V/III | Growth rate [µm/h] | MBL nominal structure | RMS* [nm] | Defect density† [$cm^{-2}$] |
|---|---|---|---|---|---|---|---|---|
| 1b | A1345 | (100) ± 0.02° | 650<br>ramp 650 to 620<br>620 | 130<br>130<br>130 | 1<br>1<br>1 | 400 nm $In_xGa_{1-x}As$ grading ~0<x<23 (initial part of a full parabola designed to reach 33%In in 1 µm)<br>500 nm $In_xGa_{1-x}As$ linear grading 23<x<45<br>190 nm $In_xGa_{1-x}As$ linear grading 45<x<53 | 6 | - |
| 1c d | A0884 | (100) ± 0.02° | 650<br>ramp 650 to 620<br>620<br>620 | 130<br>130<br>130<br>520 | 1<br>1<br>1<br>0.5 | 400 nm $In_xGa_{1-x}As$ grading ~0<x<23 (initial part of a full parabola designed to reach 33%In in 1 µm)<br>500 nm $In_xGa_{1-x}As$ linear grading 23<x<45<br>190 nm $In_xGa_{1-x}As$ linear grading 45<x<53<br>300 nm InP cap | 3 | $1\times10^7$ |
| 1 e f | A1467 | (100) ± 0.02° | 650<br>ramp 650 to 620<br>620<br>600<br>730 | 130<br>130<br>130<br>120<br>180 | 1<br>1<br>1<br>1<br>0.7 | 400 nm $In_xGa_{1-x}As$ grading ~0<x<23 (initial part of a full parabola designed to reach 33%In in 1 µm)<br>500 nm $In_xGa_{1-x}As$ linear grading 23<x<45<br>190 nm $In_xGa_{1-x}As$ linear grading 45<x<53<br>50 nm $Al_{0.48}In_{0.52}As$<br>300 nm InP cap | 2 | $2\times10^7$ |
| 1 g h | A1359 | (100) ± 0.02° | 650<br>ramp 650 to 620<br>ramp 620 to 630<br>630 | 130<br>130<br>450<br>450 | 1<br>1<br>1<br>1.14 | 400 nm $In_xGa_{1-x}As$ grading ~0<x<23 (initial part of a full parabola designed to reach 33%In in 1 µm)<br>500 nm $In_xGa_{1-x}As$ linear grading 23<x<45<br>190 nm $In_yGa_{1-y}P$ linear grading 90<y<100<br>100 nm InP cap | 7 | $2\times10^6$ |
| 2 | A1482 | (100) ± 0.02° | 650 | 130 | 1 | 1000 nm $In_xGa_{1-x}As$ grading ~0<x<30 — "typical" part | 6 | $5\times10^5$ |
| | | | | | | "odd" part | 6 | $2\times10^6$ |
| 3 a | A0850 | (100) ± 0.02° | 635 | 130<br>130<br>130 | 1<br>1<br>1 | 50 nm $In_{0.48}Ga_{0.53}As$<br>800 nm $In_xGa_{1-x}As$ grading 53<x<76<br>100 nm $In_{0.6}Ga_{0.4}As$ | 6 | $2\times10^7$ |
| 3 b | A0989 | (100) + 0.4° tow. [111]B ± 0.02° | 620<br>620<br>ramp 620 to 560<br>ramp 560 to 540 | 130<br>130<br>130<br>130 | 1<br>1<br>1<br>1 | 50 nm $In_{0.48}Ga_{0.53}As$<br>400 nm $In_xGa_{1-x}As$ grading 53<x<70 (initial part of a full parabola designed to reach 76%In in 800 nm)<br>1000 nm $In_xGa_{1-x}As$ linear grading 70<x<95<br>250 nm $In_xGa_{1-x}As$ linear grading 95<x<100 | 6 | $5\times10^7$ |
| 4 a | A0946 | (100) + 0.4° tow. [111]B ± 0.02° | 620<br>620<br>ramp 620 to 560<br>ramp 560 to 540<br>540 | 130<br>130<br>130<br>130<br>130 | 1<br>1<br>1<br>1<br>0.5 | 50 nm $In_{0.48}Ga_{0.53}As$<br>400 nm $In_xGa_{1-x}As$ grading 53<x<70 (initial part of a full parabola designed to reach 76%In in 800 nm)<br>1000 nm $In_xGa_{1-x}As$ linear grading 70<x<95<br>250 nm $In_xGa_{1-x}As$ linear grading 95<x<100<br>300 nm InAs | 4 | $5\times10^5$ |
| 4 b c | A0858 | (100) ± 0.02° | 650<br>650<br>650 | 130<br>130<br>130 | 1<br>1<br>1 | 1000nm $In_xGa_{1-x}As$ parabolic grading ~0<x<0.33<br>1140nm $In_{0.16}Ga_{0.84}As$<br>1000nm $In_xGa_{1-x}As$ parabolic grading 23<x<0.53 | 15 | - |

* Calculated from 10x10µm$^2$ AFM images after standardised flattening, averaged out after several images collected per sample

† Calculated from 10x10µm$^2$ or 50x50µm$^2$ AFM images averaged out after several images collected per sample

## Results and discussion

In Fig. 1 we present an overview of several different metamorphic buffer layers (MBLs) designed to breach the gap between GaAs and InP (Table 1 lists growth, design details and the indium distribution presented in Fig. 1a)). The “bare” reference $In_xGa_{x-1}As$ grading has significant amount of pits visible on the top surface (sample A1345, Fig. 1b). By design for the first 400 nm the grading followed a parabolic exchange In-Ga curve to maximise defect formation at the steep part, then the moderately defected area was prolonged by a linear grading, with the slope slightly reduced towards the end, as sketched in Fig. 1a). The reference sample was subsequently planarised by several different cap choices, as shown in Fig. 1d), f) and h) (the capping was done either as a subsequent regrowth on sample A1345, or as a full separate run, without any noticeable differences between the two methods). The capping resulted in a visible improvement of the morphology and reduction of the root-mean square (RMS) roughness (from 6 nm to 2 nm, see Table 1). However, the surface roughness is not the only factor that needs to be considered with process optimisation, as step bunching or long range step organisation might not have necessarily a severe detrimental effect on a device performance, despite their contribution to higher RMS. Threading dislocations are of major relevance as well, as they are affecting the subsequent overgrowth, and their presence does not necessarily correlate with surface step organiation.

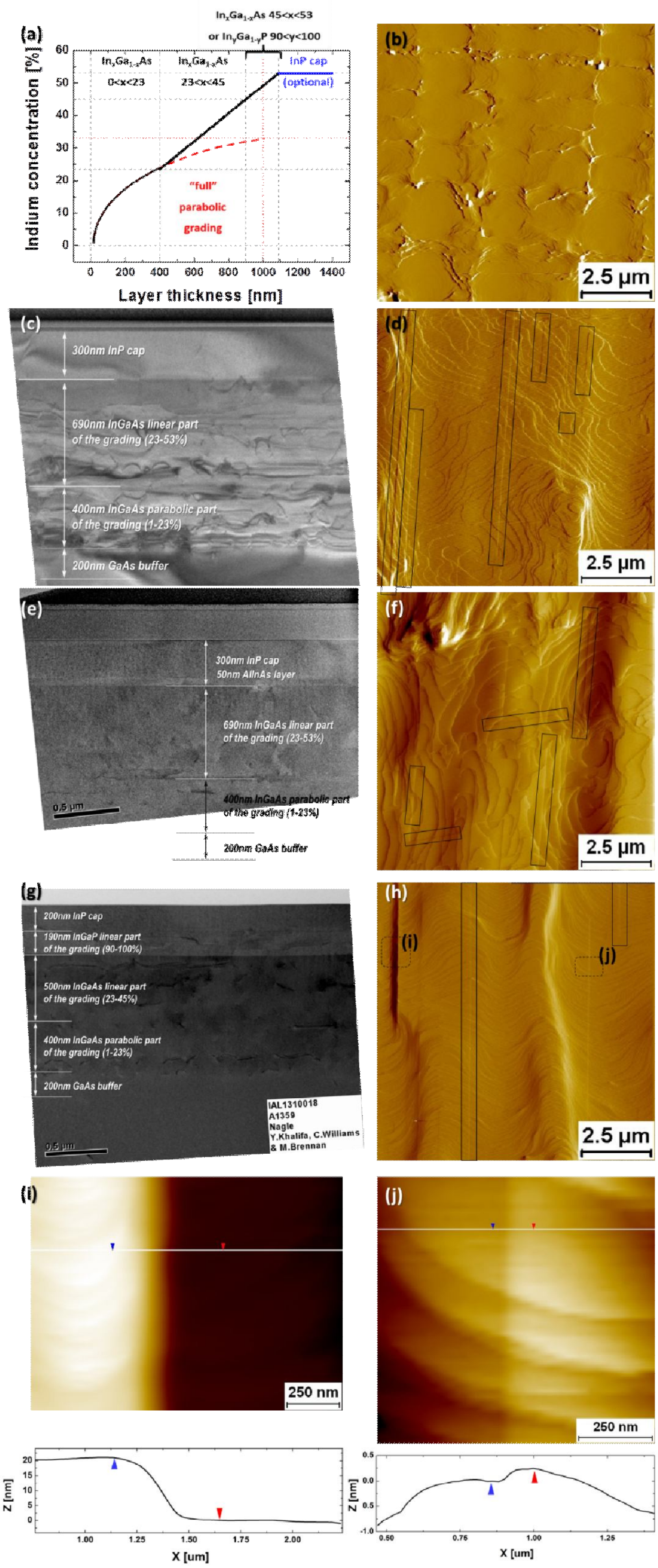

Fig. 1 (color online) Overview of investigated samples. a) simplified graph showing the design of the MBLs ; b) AFM image (signal amplitude) of sample A1345; c) and d) sample A0884; e) and f) sample A0884 ; g) and h) sample A1359; TEMs and AFM signal amplitudes, respectively. i) and j) images show zoom in (AFM signal height) to the indicated places on h) (sample A1359) and corresponding height profiles. Exemplary defect lines are marked, to guide the reader, but not all of them to avoid obscuring the image. Dashed boxes on h) indicate the zoomed in areas. For design details refer to Table 1.

Cross sectional TEM measurements are often used to estimate the density of defects. Nevertheless TEM does not always allow a precise estimation of surface defect density, as it is limited in area sampling, and a relatively low density of defects can go unnoticed. In the following we will discuss the relevance of how AFM can be applied to solve this issue.

From the AFM scans reported in Fig. 1, one can clearly see two different types of surface features: a) crystallographic steps with their bunching and meandering and b) sharp, (nearly) perpendicular lines, sometimes forming quadrilateral structures (some are highlighted), which we identify as possible dislocations threading to the surface plane, in analogy to what was proven in Ref. 9. Sometimes the distinction between the defect and crystallographic step lines in the AFM images was nontrivial, as the height of the features often corresponds to one monoatomic step and could be easily confused with a single monolayer edge. We based our assessment on our experience with the material systems described: the “normal” step organisation in InGaAs or InP layers shows step edges which are not ideally sharp, often are bunching or meandering and do have a specific (for given sample) overall trend. Defect lines observed here were unusually straight, recognisably parallel to each other, but not to normal steps, often breaking or disappearing visually for a short length and then visualising again as a prolongation of a previous part or perturbing the normal step organisation of the surface. In several cases they were going in “odd” directions, making them stand out from the overall step morphology. It should be noted that the presence of those defects can also affect the generic step organisation, aligning sometimes the epitaxial steps along the defect direction and “hiding” the defect in the step edge line, as will be discussed further in the text.

Quite clearly here AFM shows features that simple cross section TEM has a very low probability of detecting. Indeed, despite the fact that the TEM analyses were oriented as much as possible on identification of defects[15] most of them do not show any visible dislocation threading to the sample surface (Fig. 1c), d), e)). An indication of a relatively low density, at least for TEM analysis.

The calculated defect density based on the AFM imaging technique reported in Fig. 1 is here in the range of $10^6$-$10^7$ $cm^{-2}$, depending on the selected samples (see Table 1 for the reported ones). Of course, defect counting is partially arbitrary here, as it is difficult to distinguish, e.g. the feature is one long line partially interrupted (visually) by overlap with a crystallographic step or the ones in focus are two different defects. Nevertheless, TEM suffers from even greater limitation. We also noticed that the angle between the oppositely running lines can change, depending on the stack design, which would be a useful hint in further defect identification work.

The defects probably originate at the buried interfaces (typically interfacial misfit dislocations) and propagate (evolving and interacting with each other) in the bulk towards the surface[16,17], where they take the form of steps ranging in height from ~3Å to significant values (>10nm), when strong step bunching overlaps with the defect line. (Fig. 1 i) and j))

The method is not specific to the InP based samples analysed and can be applied also to other III-V materials and alloys. The only requirement is an otherwise well-ordered surface organisation and moderate amount of significant pits and 3D growth on the surface. In Fig. 2 we present an $In_xGa_{1-x}As$ MBL, which showed a visibly worse quality spot on the sample surface (Fig. 2 a)), most probably due to a particle fallen on the substrate during the early stages of the growth. Despite the same RMS value, AFM scans allow a quick identification of the two regions by showing much higher defect density in the “odd” part. To enhance the visibility of the lines in a single AFM scan, which in many cases were perfectly perpendicular to each other, we abandoned the good practise of scanning the samples across the main visible surface pattern and turned the scanning direction by 45º (Fig. 2 d). This (i.e. angled scanning) slightly disrupts the measurements of the exact epitaxial steps height, but allows for clearer imaging of defect lines running in different directions. The scans are also easily scalable, to allow for defect counting over a large area within one image (this is why we do not routinely present here the large area scans – they were collected for the purpose of quality assessment for all the samples, but require high magnification and good resolution to be usable for defect counting purposes). Nevertheless the significant residual strain in the structures (e.g. parallel strain in the top layer of sample A1482 was $\varepsilon_p$=-0.0049%) leads to long range organisation of the sample surface steps (the surface of sample appears “wavy”), therefore the lines running along the “waves” might be hiding a threading dislocation or be completely free of it, without any AFM signature. Indeed, this reduces the accuracy of our method, and should be kept in mind.

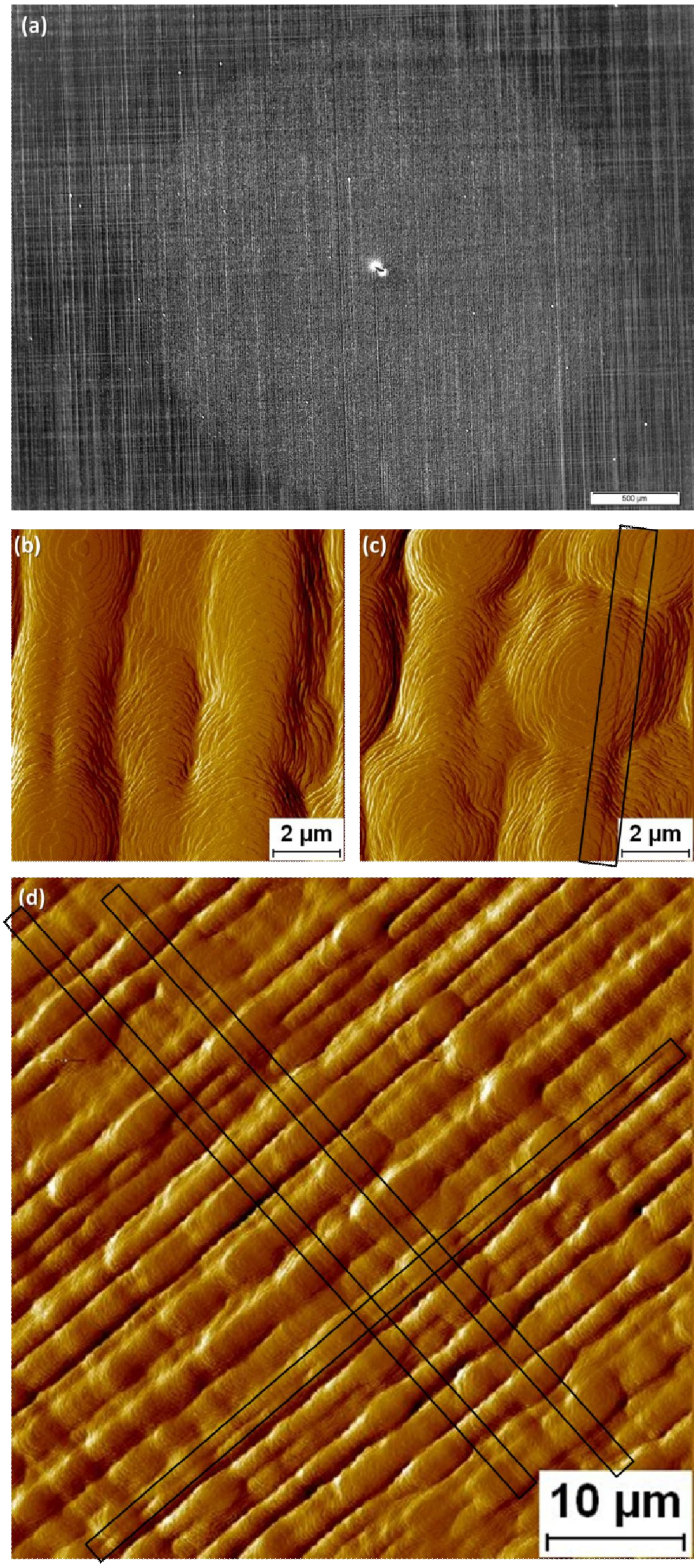

Fig. 2 (color online) Sample A1482 a) N-ICM image showing the defected spot; AFM (signal amplitude) of the sample collected in b) a « typical » area, c) and d) inside the « odd » region. Exemplary defect lines are marked, to guide the reader, but not all of them to avoid obscuring the image. For design details refer to Table 1.

As a last example, the same technique was used during optimisation of an MBL design, aimed at bridging the lattice constant between InP and InAs. In Fig. 3 we show a partial and full grading originating from the InP lattice parameter (the final, optimised design with InAs overgrowth was actually presented in Ref. 4).

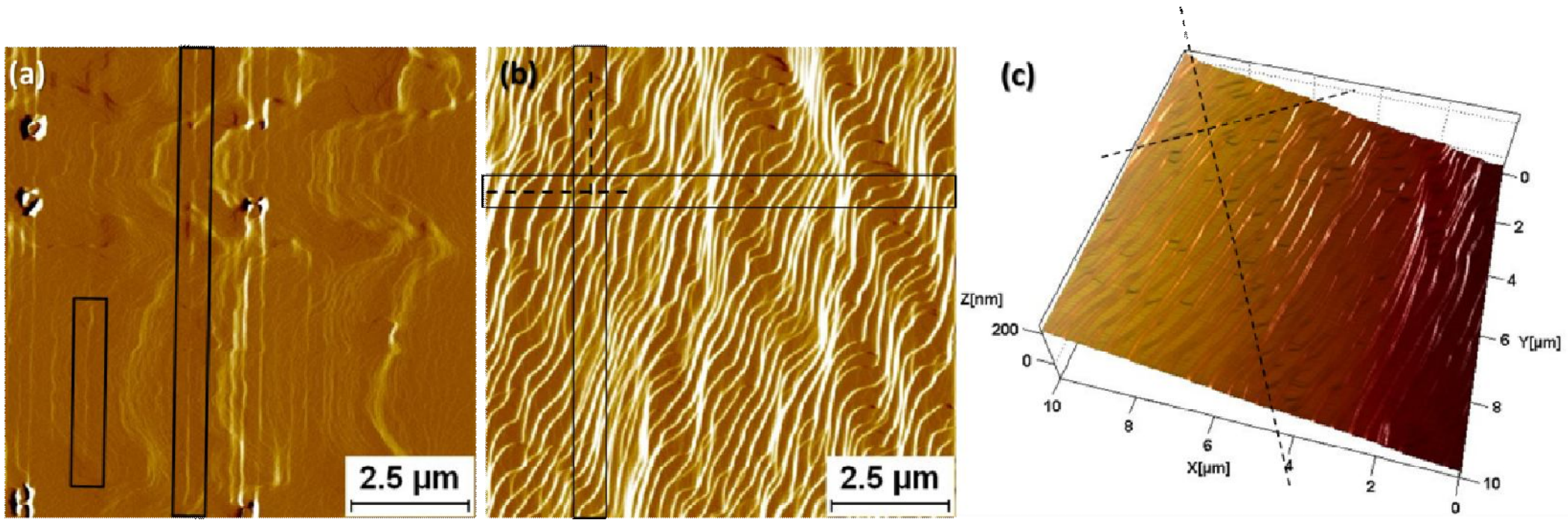

**Fig. 3 (color online) AFM images (signal amplitude) of samples a) A0850, b) A0989, c) 3D height reconstruction of A0989. Exemplary defect lines (dash in case of the defects hidden in step organisation) are marked, to guide the reader, but not all of them to avoid obscuring the image. For design details refer to Table 1.**

In Fig. 3 a) the defect lines are obvious; however they can be also identified in Fig. 3 b), but appear hidden in the step structure. That should be carefully taken into consideration to avoid recognising as good samples structures what indeed shows/hides a substantial defect density. Here, the threading dislocation actually shows up as an unnatural and unusual step organisation: subsequent steps are aligning along the same line, regardless of their original growth direction and subsequent steps have kinks lining up in the opposite direction. This becomes more evident in Fig. 3 c), with a 3D height reconstruction of that same scan, but nevertheless it takes careful surface analyses and a lot of experience to recognise that there is actually anything out of ordinary with the morphology of that sample.

The purpose of this work is not to show a best sample design, but to present a useful optimisation technique for an everyday laboratory work. However we would like to stress that the evaluation method leads to obtaining successful growth results, and it is indeed possible with the MBL design to obtain defectivities below the $10^6$ $cm^{-2}$ level. For example in Fig. 4 a) we present a scan of sample A0946 in which, by capping, we could reduce the number of the visible defect lines (estimated with the imaging limitations already discussed), suggesting a low defect count and preferential choice of this structure above the others for the specific purpose (to be compared with A0989 presented in Fig. 3 b) and c), where we show the same structure without the cap).

We would like to stress here the necessity for the top view AFM analysis in sample characterisation studies. Despite the well-known limitations of the TEM technique, often the literature reports restricts themselves to showing a single, cross sectional TEM as a representation of the growth quality. And, as presented in Fig. 4 b) and c), it is evident that sometimes it can clearly be not the full picture, and careful analysis is needed.

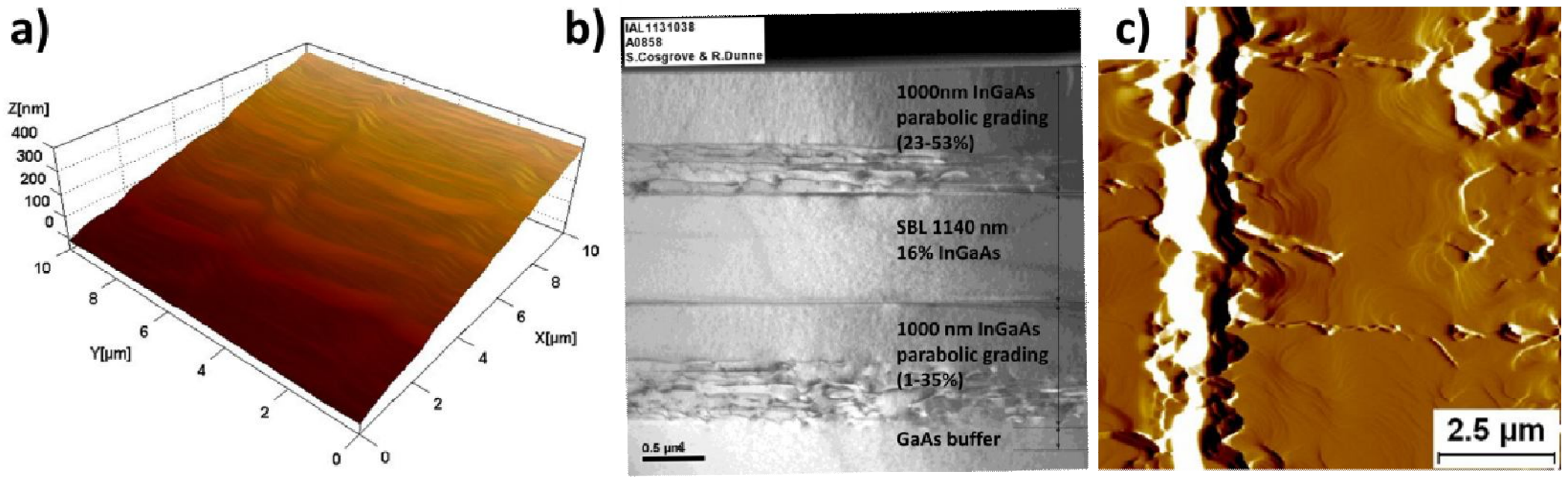

**Fig. 4 (color online) a) AFM image (3D height reconstruction) of sample A0946; b) TEM and c) AFM (signal amplitude) images of sample A0858, and, respectively. The images show big discrepancy between the "quality" that can be assessed to the given structure with two different imaging techniques. For design details refer to Table 1.**

## Conclusions

We have shown that AFM method is possible to obtain relevant information about defect density, without relying on more advanced microscopy techniques. The method has huge advantages over the other methods, e.g. fast, non-destructive and relatively cheap, allowing for quality control and direct comparison of numerous samples usually grown by MOVPE during process optimisation. It also allows for observation of two dimensional defect distribution and simultaneous measurements of defects propagating in different crystallographic directions.

## Acknowledgement

This research was enabled by the Irish Higher Education Authority Program for Research in Third Level Institutions (2007-2011) via the INSPIRE programme, by Science Foundation Ireland under grants 10/IN.1/I3000, 12/RC/2276 (IPIC) and by an EU project FP7-ICT under grant 258033 (MODE-GAP). We gratefully acknowledge Dr. K. Thomas for the MOVPE system support, Intel Corporation for partial financial support and Intel Ireland for TEM analysis (M. Brennan, S. Cosgrove, Y. Khalifa, C. Williams, R. Dunne and D. Aherne) and other support (R. Nagle).